\documentclass[nopreprintline,preprint,12pt,authoryear]{elsarticle}

\begin{document}
%\linenumbers
\begin{frontmatter}
\title{Processing System for Coherent Dedispersion of Pulsar Radio Emission}
\author[aff]{I.~A.~Girin}
\author[aff]{S.~F.~Likhachev}
\author[aff]{A.~S.~Andrianov}
\author[aff]{M.~S.~Burgin}
\author[aff]{M.~V.~Popov}
\author[aff]{A.~G.~Rudnitskiy\corref{cor}} % перенес себя в алфавитном порядке
\author[aff]{V.~A.~Soglasnov}
\author[aff]{V.~A.~Zuga}
\affiliation[aff]{organization={Astro Space Center, Lebedev Physical Institute, Russian Academy of Sciences},%Department and Organization
            addressline={Profsoyuznaya str. 84/32}, 
            city={Moscow},
            postcode={117997}, 
            country={Russian Federation}}
\cortext[cor]{Corresponding author: almax1024@gmail.com}
\begin{abstract}
The work describes a system for converting VLBI observation data using the
algorithms of coherent dedispersion and compensation of two-bit signal
sampling. Coherent dedispersion is important for processing pulsar observations
to obtain the best temporal resolution, while correction for signal sampling
makes it possible to get rid of a number of parasitic effects that interfere
with the analysis of the diffraction pattern of pulsars. A pipeline has been
established that uses the developed converter and the ASC Software Correlator,
which will allow reprocessing all archived data of Radioastron pulsar
observations and to conduct a search for giant pulses, which requires the best
temporal resolution.
\end{abstract}

%\begin{keyword}



%\end{keyword}

\end{frontmatter}

\section{Introduction} \label{sect:intro}
The Radioastron project comprised the 10-meter space radio telescope (SRT) that
together with ground VLBI antennas formed a space-ground interferometer with a
maximum baseline projection of up to 380000 km and a record angular
resolution of about 8~$\mu$as \citep{Baan2022}. The SRT was launched on the 18th
of July 2011, and successfully operated till January 2019.

The strategy of the Radioastron mission was to archive all the original raw
baseband data. Such an approach provides an ability to re-correlate the original
data if new scientific problems arise or improved methods of data reduction and
interpretation are developed.  At the end of the mission operation, the total
volume of the raw data was approximately 3500~TB.  More details about the
archive and observations database can be found in \citet{Shatskaya2020a,
  Shatskaya2020b}.

Pulsar observations were an important part of the Radioastron scientific program
\citep{Kardashev2013, Kardashev2017}. They were conducted at 324~MHz (P-band) or
1668~MHz (L-band) and in some cases at both frequencies simultaneously. The
single intermediate frequency (IF) bandwidth of the Radioastron was 16~MHz. The
P-band receiver supported one sub-band, while 1668~MHz had two 16~MHz
sub-bands. In total, the data of 25~pulsars were accumulated during the
Radioastron operation, which included 98 observations having total duration of
250~hours. Usually, at least two large ground-based radio telescopes
participated in the space-ground VLBI sessions, such as GBT, Arecibo, or the
WSRT aperture synthesis system.

As compared to single dish observations of pulsars that are usually performed
using temporal resolution, $\Delta t$, of the order of $1\,\mu{\mathrm s}$,
interferometric observations are conducted with lower values of $\Delta t$.  In
particular, for the Radioastron pulsar observations $\Delta t=62\,$ns.  High temporal
resolution of the recorded signal combined with high sensitivity of the
ground-based telescopes participating in the observations
and their ability to simultaneously measure the flux density in two polarization
channels permits, in principle, to study the phenomena that involve rapid
variations of intensity and/or polarization.  An example of such a phenomenon is
the longitudinal dependence of the polarization of giant pulses of the Crab
pulsar.  An analysis of that dependence was carried out by \citet{Main2021}, who
detected the difference between locations of emitting regions where pulses and
interpulses originate.

In studying the physics of pulsars, to take full advantage of the high temporal
resolution of Radioastron observations it is necessary to ``dedisperse'' the
received signal, that is to correct it for smearing caused by the frequency
dependence of the propagation speed of radio waves in the intervening
interstellar plasma.  The standard Radioastron data processing includes the
procedure of the so called ``incoherent'' dedispersion,  but precise
measurements of rapid intrapulse variations in the data strongly influenced by
the interstellar plasma dispersion may require the application of more
accurate and much more computer-intensive ``coherent'' dedispersion 
(see \cite{Hankins1975}, \citet{Straten2011}).

High temporal resolution of the interferometric data is achieved partly through
the low number of bits in each individual readout of the observed signal.  The
ground- and space-based observations in the Radioastron project were performed
using one- and two-bit digitizing, respectively.  Precision of observations with
a low number of digitizing levels depends critically on proper choice of  the
quantization thresholds.  The standard Radioastron data processing routines are
based on the assumption that the thresholds are set to values close to optimal
as described by \cite{Thompson2017}.  

The assumption usually holds for objects with slowly varying flux density, where
necessary adjustments are performed by an automatic gain control (AGC) system.
However, in observing pulsars the AGC is not used as it cannot function
properly when the received signal is rapidly variable.  Consequently, the
digitizing may be performed with quantization thresholds that deviate
significantly from optimal level, and reduction of the data using the standard
approach introduces additional error in the final results.  To minimize that error,
the algorithms of processing of the digitized signal should be generalized to
the case of arbitrary values of quantization thresholds. This problem was
considered by \cite{Jenet1998}. 

An example of how using a standard procedure for processing pulsar observations
leads to erroneous conclusions was presented by \cite{Popov2023}.  In that paper
it was shown that the dynamic spectra of the interstellar scintillations of the
PSR B1237+25 obtained using the standard data processing algorithms exhibit
frequency shifts that depend on the longitude and, if real, could be interpreted
as a signature of the ``interstellar interferometer'' effect.  However, the
shift almost disappears when the coherent dedispersion and proper correction
for the actual values of quantization thresholds is applied.

In this paper we describe the software for processing pulsar observations
performed by the Radioastron.  The newly developed preprocessor of the raw
observational data and the modified ASC software correlator allow one to
perform the coherent dedispersion and correctly account for digitization with
non-optimal quantization thresholds.

The paper is organized as follows: in Section \ref{sect:dedispersion_intro} we
briefly describe the methods of dedispersion, 
and the numerical method used for
correction for digitizing is outlined in Section \ref{sect:2-bit}.  Technical
details of the implementation are described in Section
\ref{sect:implementation}.  We present the results of application of the
described methods to Radioastron observations of the PSR B1237+25 in Section
\ref{sect:results} and conclude in Section \ref{sect:conclusions}.
\section{Correction of distortions caused by the dispersion}
          \label{sect:dedispersion_intro}
\label{ddisp} 
Due to the dispersion of radio waves, emission propagating through the ionized
interstellar plasma arrives at the observer with a delay depending on the signal
frequency. 
A short quasi-monochromatic impulse emitted at the low-frequency edge of the
observing band will be received later than a synchronously emitted impulse at
the high-frequency edge.  The relative delay, $\tau_{\mathrm{d}}$, is given by
\begin{equation}
  \tau_{\mathrm{d}}=D\cdot\left(\frac{1}{f_0^{2}}-\frac{1}{(f_0+B)^{2}}\right),  
  \label{eq:disp}
\end{equation}
where $f_0$ and $B$ are the lowest observing frequency and the bandwidth,
respectively.  The coefficient $D$~(s$\cdot$MHz$^{2}$) is related to the dispersion
measure $DM$~(pc$\cdot$cm$^{-3}$) by
\begin{equation}
  DM=2.410\times 10^{-4}\cdot D,
  \label{eq:rel}
\end{equation}
and the dispersion measure is defined as the integral of the electron
  column density along the line of sight to the pulsar. As a result
  of relative delay,
  the variability in integrated flux on time scales shorter than
  $\tau_{\mathrm{d}}$ is temporally smeared. 

%\bms{В нижеследующий параграф перенесено содержание фрагмента из \ref{sect:incoherent}}

There exist two methods to reduce the effect of smearing: post-detection or
incoherent dedispersion and pre-detection or phase-coherent dedispersion.
Incoherent dedispersion is performed by dividing the frequency band into many
smaller bins. The signal received in each bin is shifted in time to compensate
for the difference in arrival time, then the shifted signals are summed.
Such a method compensates for the delay between the bins, but
the dispersion within individual bins remains uncompensated. 

The coherent dedispersion is free from such a drawback. 
The method is based on the fact that the effect of dispersion on the
signal received from a pulsar can be modeled as a linear filtering operation and
the original signal can be recovered
from the received signal by performing the inverse filtering.
The effect is best described in the frequency domain. The dispersion causes
  the phase shift, 
\begin{equation}
   \phi(f)=\frac{2\pi Df^2}{f_0^2\cdot(f_0+ f)},
    \label{eq:resp}
\end{equation}
of a Fourier component of the original signal corresponding to frequency $f$.
Consequently, the Fourier transform of the dedispersed signal may be computed
by shifting the phase of each Fourier component of the observed signal by $-\phi(f)$.

In the observations of pulsars, this procedure is complicated by the fact
  that the function sought to be measured is not the Fourier spectrum of the
  signal in the strict mathematical sense, but the so called dynamic spectrum.
%%  See also https://en.wikipedia.org/wiki/Short-time_Fourier_transform
  Computation of the dynamic spectrum consists in dividing the observing
  session in many non-intersecting time intervals, usually of equal duration and
  covering the whole session, and performing the Fourier analysis of the signal
  on each interval separately.  Because of the dispersion, the interval at which
  the signal is received depends not only on the moment of emission, but also on
  the frequency. For this reason, coherent dedispersion can not be performed
  independently on each separate interval.  An approach to overcome this
  difficulty was proposed by 
\cite{Hankins1975}, who presented an elaborate description of the coherent
dedispersion technique that we follow in our study.  
\section{Correction for 2-bit Sampling} \label{sect:2-bit}
Usually, ground telescopes record the signal using 2-bit (four-level) digitizing. Four levels prescribed for such digitizing are -3, -1, +1, and +3. 

The transition threshold between the levels $1$ and $3$ is supposed to be close to the value of the original analog signal RMS ($\sigma$). The optimal value for the threshold is $t=0.9674\sigma$ in the case of 2-bit sampling \citep{Thompson2017}. %The automatic gain control (AGC) system has to support the correct values of the threshold during the observation.

Besides, it is important to switch off the telescope's automatic gain control (AGC) system in pulsar observations because the system inertia would not operate properly at the ON-pulse and OFF-pulse stages of pulsar observation. Hence a pulsar signal is an example of a non-stationary noise process. Thus, records for such signals must be corrected for 2-bit digitizing. The problem was considered by \cite{Jenet1998}. They have demonstrated that digitizing the signal before removing the dispersive effects generates unwanted systematic artifacts in the data. Namely, the ``negative'' dips appear around the pulse in the average pulse profile. 

We follow the method described in \citet{Jenet1998}. First, it is necessary to estimate the undigitized power level $\sigma_0$ using a fraction of samples $\Phi$ with values of $\pm$1 in the given portion of the signal record:

\begin{equation}
  \Phi(t)=\frac{1}{\sqrt{2\pi}\sigma}\int \limits_0^t \exp{\left({-{\frac{x^2}{2\sigma ^2}}}\right)}dx = {\rm erf} \left(\frac{t}{\sqrt 2\sigma}\right)
  \label{eq:dig}
\end{equation}

\begin{figure*}[ht]
\begin{center}
\includegraphics[angle=270,width=67mm]{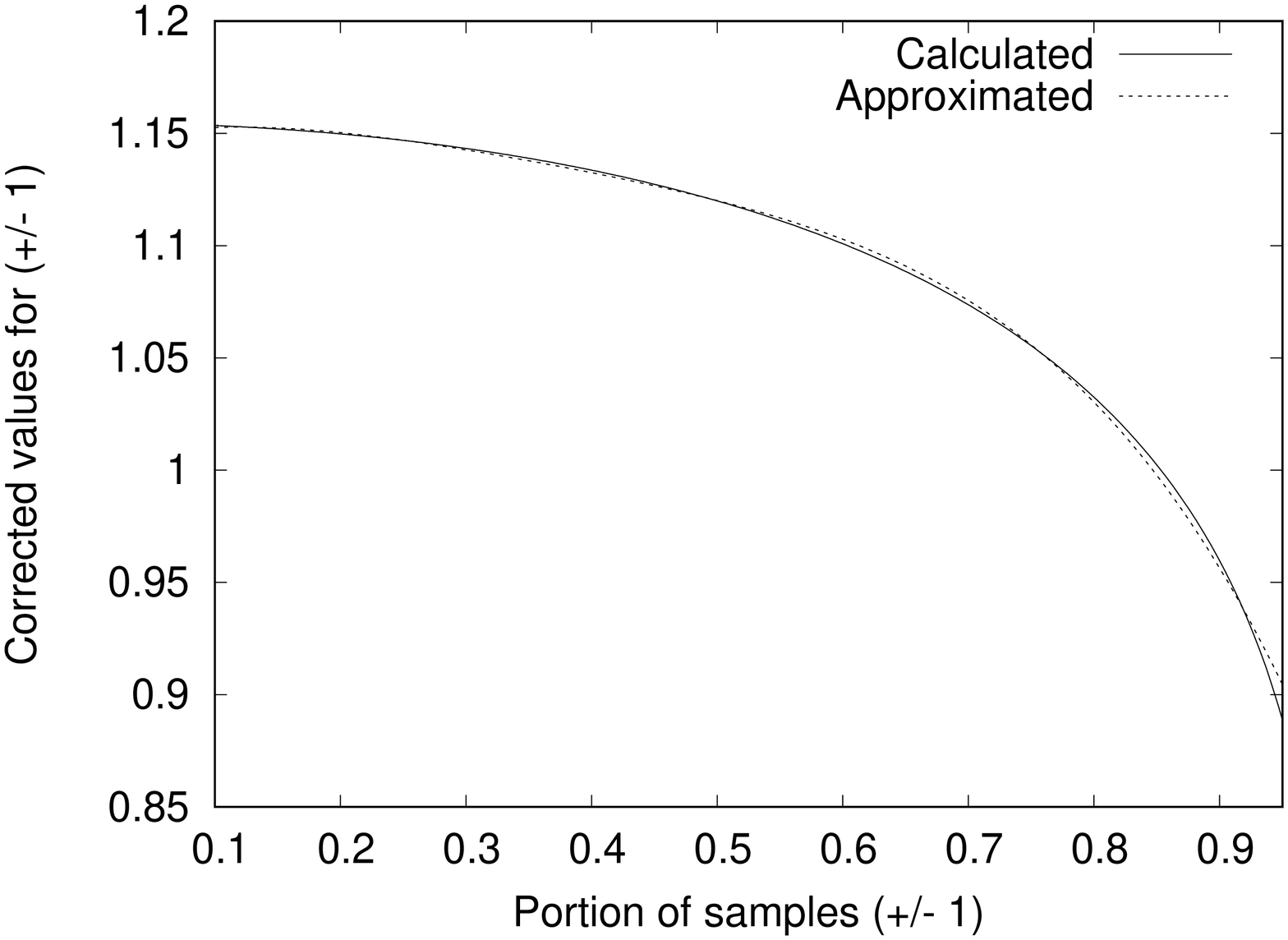}
\includegraphics[angle=270,width=67mm]{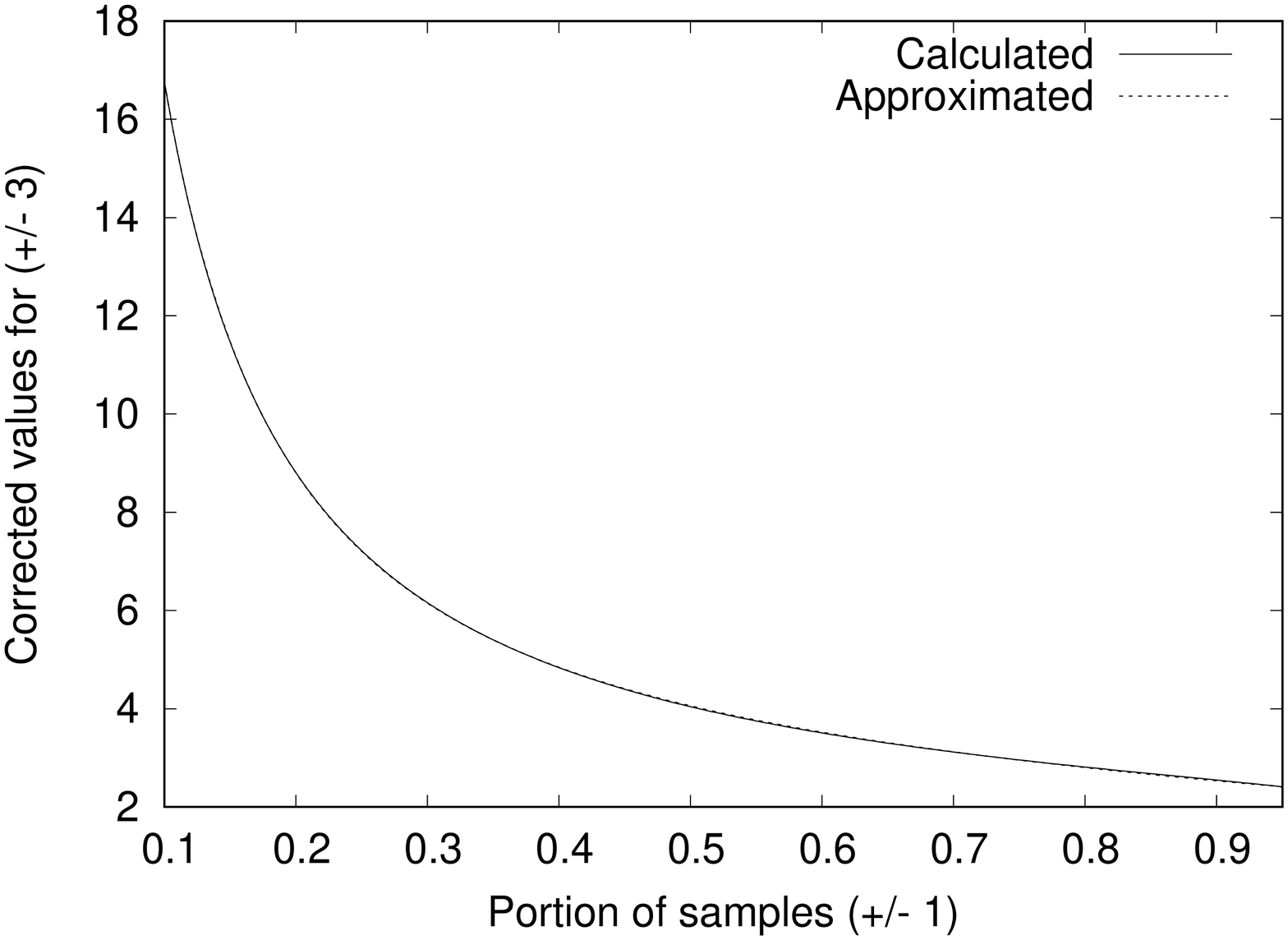}
\caption{The corrected values $y1$ and $y3$ versus observed fraction of samples equal $\pm 1$.}\label{fig:2bits}
\end{center}
\label{fig:cor}
\end{figure*}

With the value of $\sigma_0$ known it is possible to calculate the corrected values
of signal $y1$ and $y3$ instead of $\pm$1 and $\pm$3 using Eq.~(40) and (41)
from \cite{Jenet1998}. In practice, we calculated all values using $\sigma_0$ as an
independent variable, and then we consider the results with independent variable
$\Phi$. These results are shown in Fig.~\ref{fig:2bits}. To make corrected values
closer to VLBI standard, the obtained $y1$ and $y3$ were multiplied by a factor
of 2. Let us define the calculated threshold value from Eq. (\ref{eq:dig}) as
$\Lambda$.  For example, $y1(0.67)=1.085$ and $y3(0.67)=3.23$ and $\Phi=0.67$ for
$t=0.9674$ is an optimal threshold.

Approximation functions were introduced to calculate $y1$ and $y3$ from the measured fraction of samples $\Phi$ with digitized values as $\pm1$:

\begin{equation}
\label{eq:approx1}
 y1(\Lambda)=a1+b1\cdot\Lambda+c1\cdot\Lambda^2+d1\cdot\Lambda^3+e1\cdot\Lambda^4
\end{equation}
\begin{equation}
\label{eq:approx2}
 y3(\Lambda)=a3\cdot \exp{\left(-\frac{\Lambda^{b3}}{d3}\right)}+c3+e3\cdot\Lambda
\end{equation}

The polynomial approximation is sufficient for $y1$, while 
an approximation by a combination of an exponent and linear term is more suitable for $y3$.
The coefficients used in approximating functions are the following:\\
\begin{center}
$a1=1.1438(5)$, $b1=0.169(7)$, $c1=-0.96(3)$, $d1=1.62(4)$, $e1=-1.13(2)$;\\
$a3=1920(200)$, $b3=0.228(5)$, $c3=3.24(4)$, $d3=0.119(1)$, $e3=-1.39(3)$.\\
\end{center}
RMS uncertainties of the coefficients are given in brackets.
The difference between the calculated and approximated values can't be distinguished well due to the coarse scale in the figures, .
In fact, RMS residuals are 0.0012 for $y1$ approximation and 0.011 for $y3$.

\section{Implementation of Coherent Dedispersion}\label{sect:implementation}
\subsection{ASC Software Correlator}
The VLBI data are processed using a correlator. There are two types of correlators: software and hardware, which can be XF and FX correlators. Software correlators perform data processing on conventional computers or on computing clusters running operating systems such as Microsoft Windows or Linux. In turn, hardware correlators are implemented using programmable logic integrated circuits (FPGAs). Correlators of the XF and FX types differ in the sequence of operations performed, where X stands for multiplication and F for Fourier transform. Accordingly, in XF correlators, the signals are first multiplied, and then the Fourier transform is performed. In FX correlators, the Fourier transform performed and the result is multiplied.

A dedicated software correlator (the ASC correlator) was developed in Astro Space Center to process the data of the Radioastron mission. It is a software FX correlator that calculates auto and cross spectra for all possible combinations of baselines and polarizations. The detailed description is provided in \cite{Likhachev2017}.

The ASC software correlator operates on a high-performance computing (HPC) cluster using the message-passing interface (MPI) for parallel calculations. Its total computing power is 1~Tflops. This software correlator supports several modes of data processing: continuous spectrum, spectral lines, and several types of pulsar data processing, including the search and correlation of giant pulses. It supports widespread nomenclature of VLBI baseband data formats such as MarkIV/A, Mark5B, VDIF, RDF, LBA, etc. The ASC Correlator processed about 95\% of Radioastron data, including the pulsar one. The complexity of the pulsar VLBI data processing primarily lies in the need to compensate for the dispersion of the signal.

\subsection{Compensation of Dispersion}\label{sect:compensation}
\subsubsection{Incoherent Dedispersion} \label{sect:incoherent}

Incoherent dedispersion implemented in the ASC software correlator is performed in several steps. The pulsar period is divided into bins. The correlator calculates the spectra for each bin. The delay of signal in each bin then is compensated using a polynomial model provided by the TEMPO2 software package \citep{Edwards2006}. Further, each bin is averaged over the observation time. As a result, each bin will contain the signal with removed dispersion. The dispersion effect will decrease linearly with the increasing number of bins. Accordingly, it is necessary to sum only the data that contain fringes of pulses and discard the rest. It is important to compensate the pulse shape with a dispersive delay \citep{Likhachev2017}.

\subsubsection{Coherent Dedispersion} \label{coherent}
Up to now, the ASC correlator has not been able to perform data processing using coherent dedispersion. For these purposes we have developed a data converter that coherently removes the dispersion effects. This converter creates the data in a new complex format, which is supported by the updated version of the correlator. The converted data then is processed in the standard mode but with the dedispersion switched off. Such an approach minimized the changes in the ASC software correlator structure. Based on the converter, a pipeline was developed for processing pulsar observations. The operation steps of this pipeline are the following:

At first, one has to calculate the discrete Fourier transform of the recorded
signal.  Next, the result of the Fourier transform is multiplied by the complex
frequency response function given by Eq.~(\ref{eq:resp}). Inverse Fourier
transform will yield the recovered signal. It is equivalent to a convolution of
impulse response function with the recorded voltage signal in a time domain. The
discrete Fourier transform is represented with the periodical assumption. Thus,
the first $n$ points corresponding to the time interval of $\tau_{\mathrm{d}}$ will
be erroneously multiplied with the data wrapped around from the end of the
segment and it must be discarded. So in this approach the first portion of the
signal will be lost in the interval $\tau_{d}$.

\begin{figure*}[htb]
\begin{center}
\includegraphics[angle=0,width=90mm]{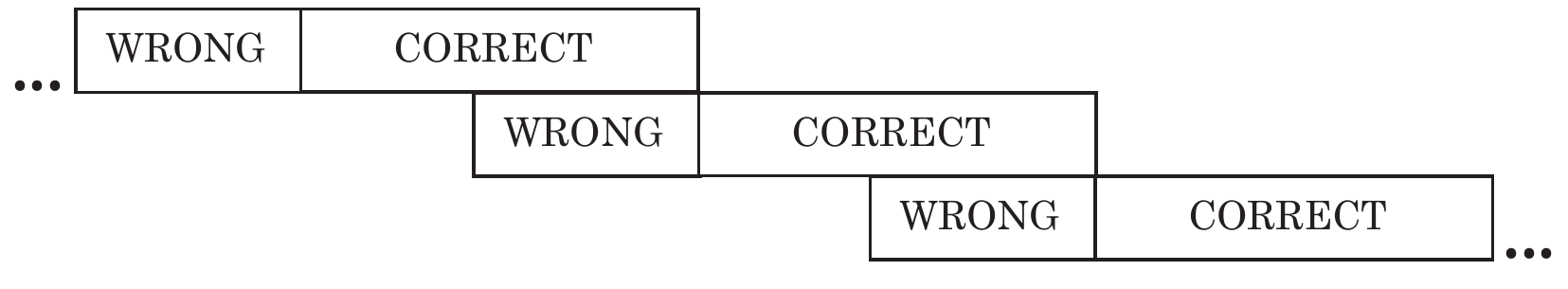}
\caption{The scheme of sequential sampling of recording sections with a duration $T$ from a continuous record.}\label{fig:diagram}
\end{center}
\end{figure*}

Fig.~\ref{fig:diagram} illustrates the scheme of sequential sampling of signal
sections of duration $T$ from a continuous recording. Sequential values of
starting time $t_i$ are determined by the relation
$t_i=t_0+i(T-\tau_{\mathrm{d}})$. It is clear that $T$ must be larger than
$\tau_{\mathrm{d}}$ and the efficiency of data reduction is
$1-\tau_{\mathrm{d}}/T$. More details on data reduction approach used in our
processing system are given below in Section~\ref{sect:implementation}.

\begin{itemize}
    \item Read raw baseband data sample of $M$ size. Supported data formats are the same as for ASC Correlator: RDF (Radioastron Data Format), MarkIV/A, Mark5B, K5A, LBA, and VDIF.
    \item Split the sample $M$ into $P$ pieces, each piece of $N$ size ($P=M/N$). 
    \item Determine a parameter $\Lambda$ by counting portion of samples equal to $\pm$1 and  calculate the values $y1$ and $y3$ for each data sample in the $P$th piece using equations \ref{eq:approx1},\ref{eq:approx2};
    \item Replace every sample $s_i$  by the values $y1$ or $y3$,
    thus  performing 2-bit sampling correction;
    \item Calculate the matrix $\phi_{i}$ of the phase shifts depending on the frequency $f_{i}$: 
        \begin{equation}
        \phi_{i}= \frac{2\pi D f_i^2}{f_0^2\cdot(f_0+f_i)}    
        \label{eq:matrixshifts}
        \end{equation}
    \item Perform a complex $M$-point Fourier transform for each data chunk numbered $k=1,...,P$;
    \item Multiply Fourier spectra by a correction complex function $R_i=e^{-\phi_{i}}$;
    \item Add $M$ more zeros to the end of the array to get an array of $2 M$ values;
    \item Perform inverse FFT to obtain complex $M$ samples;
    \item Write the converted data to the output file in half-precision float (2~bytes) format.
\end{itemize}

The converter is capable to adjust the parameters of the FFT size, and the data sample size. At the output a new data file is generated in the special format, which can be used in the ASC correlator.

Bit statistics (the number of $\pm1$ and $\pm3$) are estimated with a floating average method for $P$ data piece size ($N$). The statistics are calculated for each data point in the interval [current position - N/2, current position + N/2]. Further, the 2-bit sampling correction coefficient is considered by a function of estimated statistics (\ref{eq:approx1} and \ref{eq:approx2}).

The correlated data will be in a UVX format where cross and auto spectra are represented as a float (4 bytes number format). The UVX files can be converted straight to the widespread IDI-FITS format for further analysis. Finally, bandpass correction and noise cleaning can be applied to the correlated data using the corresponding calibration tools of ASL software package \cite{Likhachev2020}.

\subsubsection{Data Format}\label{sect:format}
The developed converter uses its special format as an output (No Packet Data or NPD), which is supported by the ASC software correlator. The first 512 bytes are dedicated to the header, which is a string that contains the information about the observation parameters separated by commas and includes: version, date (DD/MM/YYYY), time (HH:MM:SS.sss), bandwidth ($1\cdot10^{-6}\times 1/dt$ -- full bandwidth including all sub-bands), number of channels ($NCH$, polarization channels and sub-band channels), number of bits per sample. 

The header is followed by half-precision float FP16$\times$NCH data. The single-time data sample is (2$\times$NCH) bytes. The order of channels is the same as for the initial input raw baseband data. At the same time, the lower sub-band is converted to the upper one, which in turn requires a shift in the reference frequency for it when correlating the converted data.

\section{Results} \label{sect:results}

We have performed several data processing experiments with the Radioastron observations of the pulsar B1237+25 to test the effectiveness of the developed coherent dedispersion data converter. The observations were conducted at 324~MHz with a 16~MHz IF bandwidth. Arecibo and Green Bank radio telescopes took part in the session as ground support. The dispersion measure of the pulsar is $9.3$~pc$\cdot$cm$^{-3}$. The estimated time smearing is 36.35~ms in the band of 316-332~MHz. For this test, we used the following parameters of the number of samples and pieces: $M=10^{6}$ and $P=10^{3}$.

\begin{figure*}[ht]
\begin{center}
\includegraphics[angle=270,width=67mm]{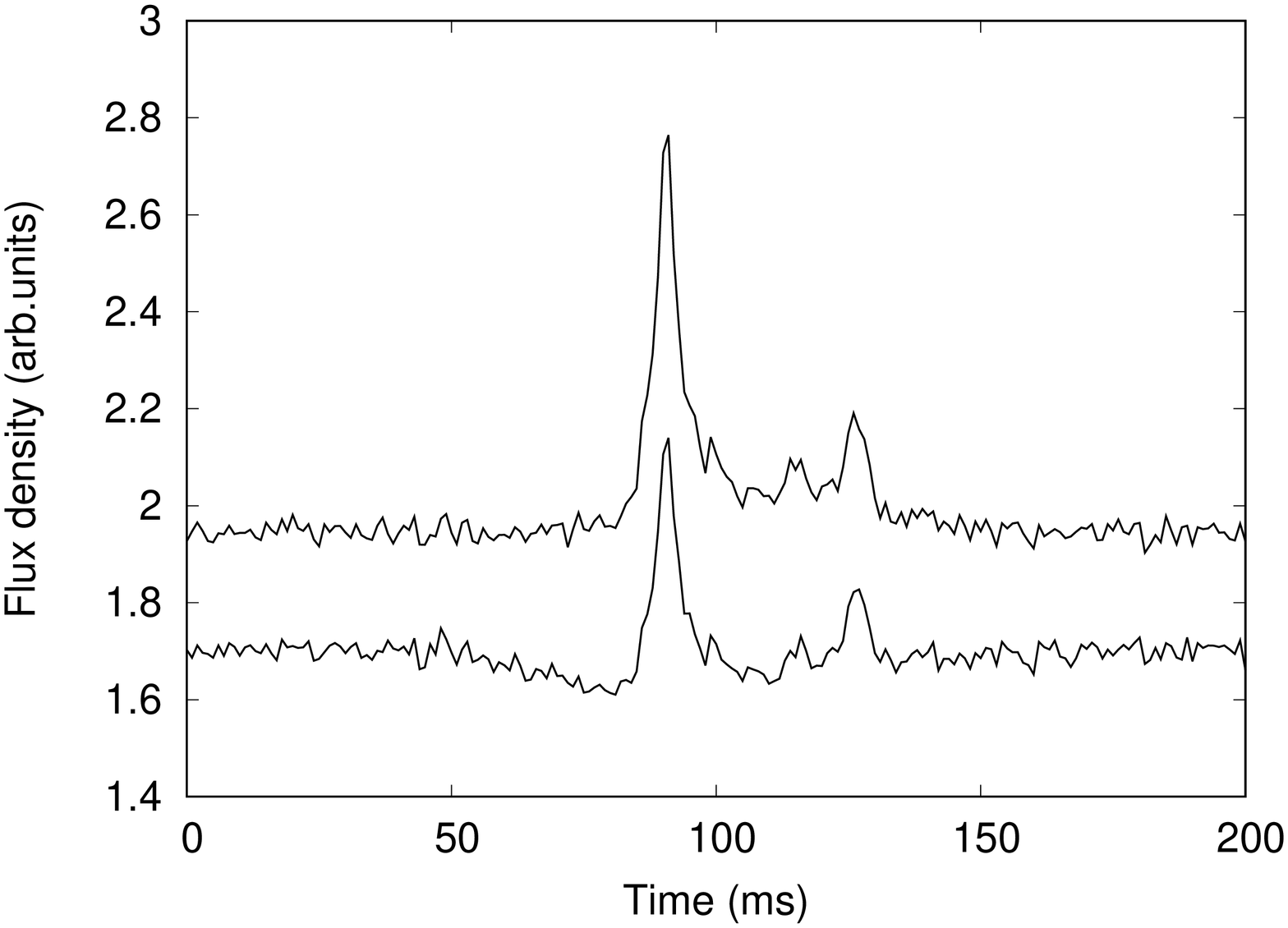}
\includegraphics[angle=270,width=67mm]{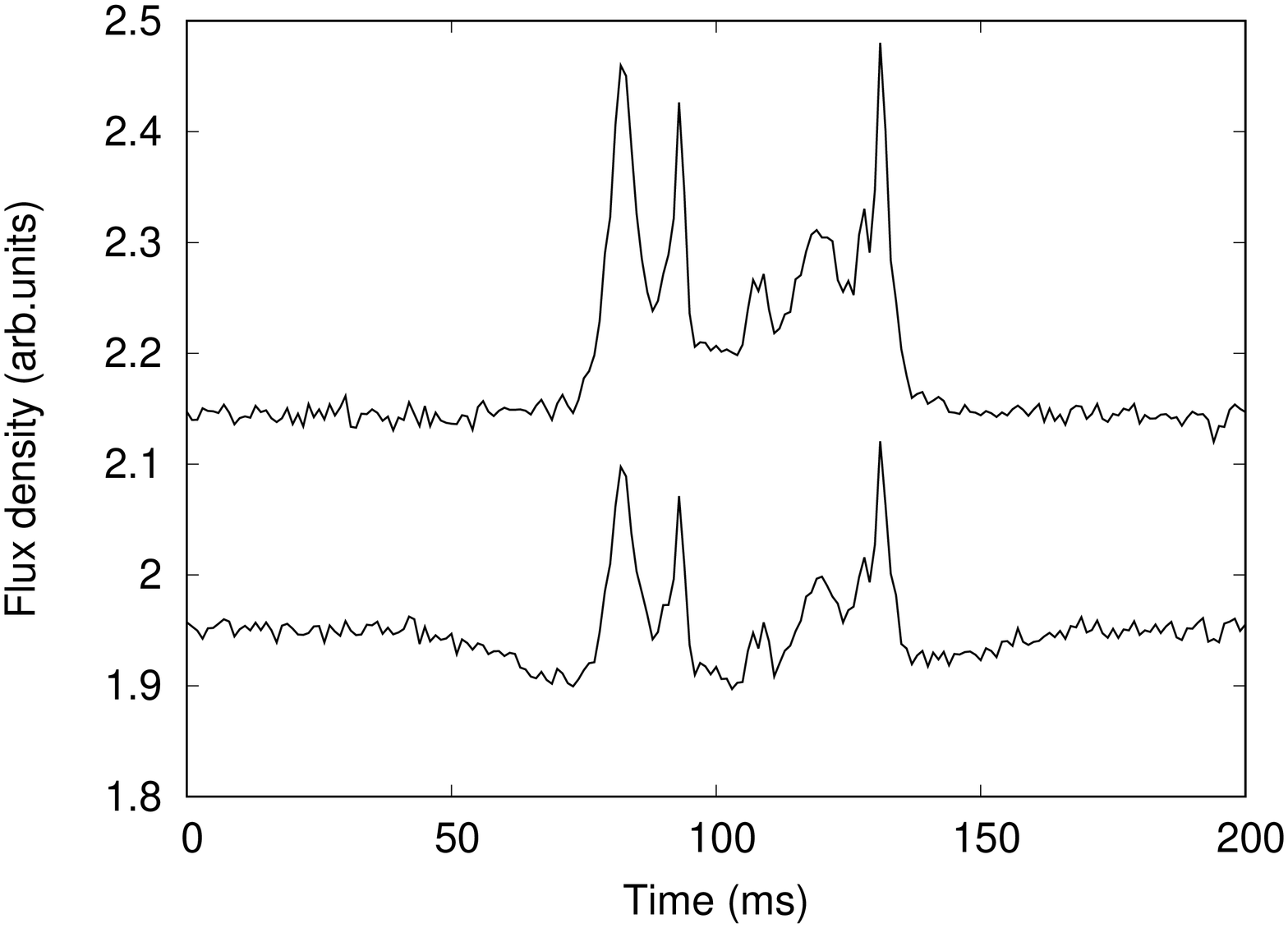}
\caption{Comparison of pulsar signals for B1237+25 restored without correction for 2-bit sampling (lower curves), an with such correction applied (upper curves). The left panel shows a single pulse, while the right panel presents a pulse averaged over 10~seconds (7~pulses). Time resolution is one millisecond.}
\label{fig:prof}
\end{center}
\end{figure*}

Fig.~\ref{fig:prof} (left panel) shows the reconstructed signal smoothed by 1~ms for one individual pulse recorded at the Green Bank radio telescope. The lower curve shows the signal after coherent dedispersion, but without 2-bit sampling correction. The upper curve shows the signal with sampling correction applied. It can be seen that the applied 2-bit correction eliminates parasitic signal dips around the pulse and improves the SNR by about 20\%.

Fig.~\ref{fig:prof} (right panel) gives a similar illustration for a pulse averaged over 10 s. in time (7~individual pulses). The distortions caused by two-bit digitization are complex and need to be corrected.

Two-bit sampling of the signal causes additional distortion in the radio spectra of pulsar pulses. These radio spectra contain important information about the scattering and scintillation parameters of the interstellar plasma inhomogeneities, namely, the so-called diffraction pattern. From an astrophysical point of view, it is of interest to compare the diffraction pattern for different parts of the pulsar averaged profile. The detection of a shift in the diffraction pattern with longitude would indicate that the radio emission at different longitudes occurs in spatially different parts of the pulsar's magnetosphere, i.e. it would be possible to measure such localization.

\begin{figure*}[ht]
\begin{center}
\includegraphics[angle=270,width=90mm]{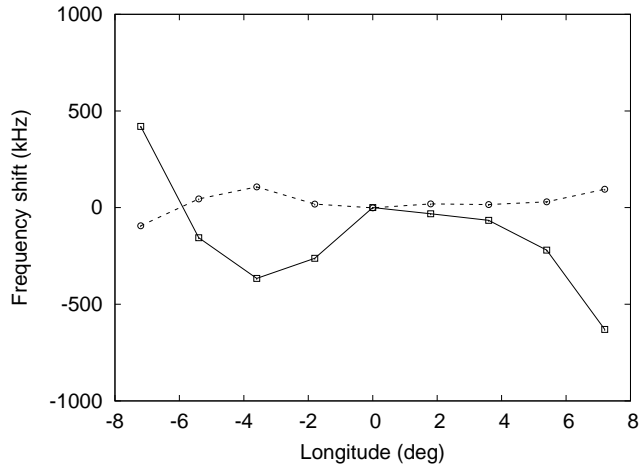}
\caption{Frequency shift between dynamic spectra obtained at different longitudes of the mean profile of B1237+25. The solid curve passing through the squares corresponds to the analysis of the spectra obtained without two-bit digitization correction, and the dotted line passing through the circles corresponds to the spectra obtained for the corrected signal.}
\label{fig:shift}
\end{center}
\end{figure*}

Fig.~\ref{fig:shift} shows measurements of the frequency shift of the diffraction pattern in the spectrum of the B1237+25 pulsar as a function of the longitude of the averaged profile. The observations of 20 minutes were carried out with the Arecibo radio telescope. The solid curve with squares corresponds to the measurements obtained without two-bit digitization correction and the dotted line with circles corresponds to the measurements for the corrected signal. The notable effect of the diffraction pattern drift with longitude disappears after the correction. This case is considered a separate study.

The program supports multi-threading and we have estimated the approximate performance on the VDIF data of the Green Bank telescope of B1237+25 which had two polarization channels (LCP and RCP) and one sub-band with a bandwidth of 16~MHz. The measured time ratio was 1:14, thus it takes 14 seconds of real-time to convert one second of such data.

\section{Conclusions}\label{sect:conclusions}
The developed converter allows not only perform of coherent dedispersion but also correct the digitized signal for two-bit sampling. This tool, together with the ASC software correlator and the ASL software package, forms a new pipeline for processing the pulsar data of the Radioastron project. This pipeline was tested on the observational data of the pulsar B1237+25 observed by the Radioastron space-ground interferometer. The corrected data then were used in the studies in \cite{Popov2023}.

Processing of pulsar data with compensation for both dispersion and 2-bit sampling is an important point and allows the removal of the parasitic effects in the data that interfere with their subsequent correct interpretation. It was shown, that the distortions caused by the 2-bit sampling can introduce interference leading to false pulsar diffraction pattern drift in longitude disappearing after bit statistic compensation.

Further, this pipeline will be used to reprocess the Radioastron pulsar raw baseband data with the maximum available temporal resolution, not smeared by dispersion, and also to process any pulsar VLBI data of modern known and supported formats.

\bibliographystyle{elsarticle-harv} 
\bibliography{biblio}

\end{document}